# Prediction of Infinite Words with Automata[*]

Tim Smith[1][2]

[1] Northeastern University
Boston, MA, USA
[2] Université Paris-Est Marne-la-Vallée
Champs-sur-Marne, France
`tim.smith@u-pem.fr`

**Abstract.** In the classic problem of sequence prediction, a predictor receives a sequence of values from an emitter and tries to guess the next value before it appears. The predictor masters the emitter if there is a point after which all of the predictor's guesses are correct. In this paper we consider the case in which the predictor is an automaton and the emitted values are drawn from a finite set; i.e., the emitted sequence is an infinite word. We examine the predictive capabilities of finite automata, pushdown automata, stack automata (a generalization of pushdown automata), and multihead finite automata. We relate our predicting automata to purely periodic words, ultimately periodic words, and multilinear words, describing novel prediction algorithms for mastering these sequences.

## 1 Introduction

One motivation for studying prediction of infinite words comes from its position as a kind of underlying "simplest case" of other prediction tasks. For example, take the problem of designing an intelligent agent, a purposeful autonomous entity able to explore and interact with its environment. At each moment, it receives data from its sensors, which it stores in its memory. We would like the agent to analyze the data it is receiving, so that it can make predictions about future data and carry out actions in the world on the basis of those predictions. That is, we would like the agent to discover the laws of nature governing its environment.

Without any constraints on the problem, this is a formidable task. The data being received by the agent might be present in multiple channels, corresponding to sight, hearing, touch, and other senses, and in each channel the data given at each instant could have a complex structure, e.g. a visual field or tactile array. The data source could be nondeterministic or probabilistic, and furthermore could be sensitive to actions taken by the agent, leading to a feedback loop between the agent and its environment. The laws governing the environment could be mathematical in nature or arise from intensive computational processing.

---

[*] This is the full version of a paper accepted for publication at CSR 2016. It contains an appendix with proofs which were sketched in the body.

A natural approach to tackling such a complex problem is to start with the easiest case. How, then, can we simplify the above scenario? First, say that instead of receiving data through multiple channels, the agent has only a single channel of data. And say that instead of the data having a complex structure like a visual field, it simply consists of a succession of symbols, and that the set of possible symbols is finite. Say that the data source is completely deterministic, and moreover that the data is not sensitive to the actions or predictions of the agent, but is simply output one symbol at a time without depending on any input.

Under these simplifying assumptions, the problem we are left with is that of predicting an infinite word. That is, the agent's environment now consists of some infinite word, which it is the agent's task to predict on the basis of the symbols it has seen so far. We hope that by exploring and making progress in this simple setting, we can develop techniques which may help with the more general prediction problems encountered in the original scenario.

### 1.1 Our contributions

In this paper, we consider the case in which the predictor in the above setting is an automaton. In our model, a predicting automaton $M$ takes as input an infinite word $\alpha$ and produces as output an infinite word $M(\alpha)$, with the restriction that for each $i \geq 1$, $M$ must output the $i$th symbol of $M(\alpha)$ before it can read beyond the $i-1$th symbol of $\alpha$. If there is an $n \geq 1$ such that for every $i \geq n$, the $i$th symbol of $M(\alpha)$ equals the $i$th symbol of $\alpha$, then we say that $M$ **masters** $\alpha$.

We consider three classes of infinite words. The first are the purely periodic words, those of the form $xxx\cdots$ for some string $x$. Next are the ultimately periodic words, those of the form $xyyy\cdots$ for strings $x, y$. Finally we consider the multilinear words [21], which consist of an initial string followed by strings that repeat in a way governed by linear polynomials, for example `abaabaaab`$\cdots$.

All of the automata we consider are deterministic automata with a one-way input tape. We first examine DFAs (deterministic finite automata), showing that no DFA predictor masters every purely periodic word. We then consider DPDAs (deterministic pushdown automata), showing that no DPDA predictor masters every purely periodic word. We next turn to DSAs (deterministic stack automata). Stack automata are a generalization of pushdown automata whose stack head, in addition to pushing and popping when at the top of the stack, can move up and down the stack in read-only mode [10]. We show that there is a DSA predictor which masters every purely periodic word, and we provide an algorithm by which it can do so.

Next, we consider multi-DFAs (multihead deterministic finite automata), finite automata with one or more input heads [13]. We show that there is a multi-DFA predictor which masters every ultimately periodic word, and we provide an algorithm by which it can do so. Finally, we consider sensing multi-DFAs, multihead DFAs extended with the ability to sense, for each pair of heads, whether those two heads are at the same position on the input tape [14]. We show that there is a sensing multi-DFA predictor which masters every multilinear word, and we provide an algorithm by which it can do so. Our results are depicted in Table 1.

| $\exists \xrightarrow{masters} \forall$ | purely periodic | ultimately periodic | multilinear |
|---|---|---|---|
| DFA | × | × | × |
| DPDA | × | × | × |
| DSA | ✓ | ? | ? |
| multi-DFA | ✓ | ✓ | ? |
| sensing multi-DFA | ✓ | ✓ | ✓ |

**Table 1:** Prediction of classes of infinite words. A checkmark means that there is a predictor in that row which masters every infinite word in that column. A cross means that this is not the case.

## 1.2   Related work

A classic survey of inductive inference, including the problem of sequence prediction, can be found in [2]. The concept of "mastering" an infinite word is a form of "learning in the limit", a concept which originates with the seminal paper of Gold [11], where it is applied to language learnability. Turing machines are considered as sequence extrapolators in [4]. An early work on prediction of periodic sequences is [20], where these sequences appear in the setting of two-player emission-prediction games. Inference of ultimately periodic sequences is treated in [15] in an "offline" setting, where the input is a finite string and the output is a description of an ultimately periodic sequence. An algorithm is presented which computes the shortest possible description of an ultimately periodic sequence when given a long enough prefix of that sequence, and can be implemented in time and space linear in the size of the input, using techniques from string matching. The algorithm works by finding the LRS (longest repeated suffix) of the input and predicting the symbol which followed that suffix on its previous occurrence.

In [18], finite-state automata are considered as predicting machines and the question of which sequences appear "random" to these machines is answered. A binary sequence is said to appear random to a predicting machine if no more than half of the predictions made of the sequence's terms by that machine are correct. Further work on this concept appears in [5]. In [9] the finite-state predictability of an infinite sequence is defined as the minimum fraction of prediction errors that can be made by an finite-state predictor, and it is proved that finite-state predictability can be obtained by an efficient prediction procedure using techniques from data compression. In [3] a random prediction method for binary sequences is given which ensures that the proportion of correct predictions approaches the frequency of the more common symbol (0 or 1) in the sequence. In [16], "inverse problems" for D0L systems are discussed (in the title and throughout the paper, the term "finite automata" refers to morphisms). These problems ask, given a word, to find a morphism and initial string which generate that word (bounds are assumed on the size of the morphism and initial string). An approach is given for solving this problem by trying different string lengths for the righthand side of the morphism until a combination is found which is compatible with the input. A genetic algorithm is described to search the space of word lengths. In [6], an evolutionary algorithm is used to search for the finite-state machine with the highest prediction ratio for a given purely periodic word, in the space of all automata with a fixed number of states. In [7], the problem of successfully predicting a single 0 in an infinite binary word being revealed sequentially to the predictor is considered; only one prediction may be made, but at a time of the predictor's choosing. Learning of languages consisting of infinite words has also been studied; see [1] for recent work.

An early and influential approach to predicting infinite sequences is that of program-size complexity [22]. Unfortunately this model is incomputable, and in [17] it is shown furthermore that some sequences can only be predicted by very complex predictors which cannot be discovered mathematically due to problems of Gödel incompleteness. [17] concludes that "perhaps the only reasonable solution would be to add additional restrictions to both the algorithms which generate the sequences to be predicted, and to the predictors." This suggestion is akin to the approach followed in the present paper, where the automata and infinite words considered are of various restricted classes. Following on from [17], in [12] the formalism of sequence prediction is extended to a competition between two agents, which is shown to be a computational resources arms race.

### 1.3 Outline of paper

The rest of the paper is organized as follows. Section 2 gives definitions for infinite words and predicting automata. Section 3 studies prediction of purely periodic and ultimately periodic words. Section 4 studies prediction of multilinear words. Section 5 gives our conclusions.

## 2 Preliminaries

### 2.1 Words

Where $X$ is a set, we denote the cardinality of $X$ by $|X|$. For a list or tuple $v$, $v[i]$ denotes the $i$th element of $v$; indexing starts at 1. An **alphabet** $A$ is a finite set of symbols. A **word** is a concatenation of symbols from $A$. We denote the set of finite words by $A^*$ and the set of infinite words by $A^\omega$. We call finite words **strings** and infinite words **streams** or $\omega$-**words**. The length of $x$ is denoted by $|x|$. We denote the empty string by $\lambda$. A **language** is a subset of $A^*$. A (symbolic) **sequence** $S$ is an element of $A^* \cup A^\omega$. A **prefix** of $S$ is a string $x$ such that $S = xS'$ for some sequence $S'$. The $i$th symbol of $S$ is denoted by $S[i]$; indexing starts at 1. For a non-empty string $x$, $x^\omega$ denotes the infinite word $xxx\cdots$. Such a word is called **purely periodic**. An infinite word of the form $xy^\omega$, where $x$ and $y$ are strings and $y \neq \lambda$, is called **ultimately periodic**. An infinite word is **multilinear** if it has the form

$$q \prod_{n \geq 0} r_1^{a_1 n + b_1} r_2^{a_2 n + b_2} \cdots r_m^{a_m n + b_m},$$

where $\prod$ denotes concatenation, $q$ is a string, $m$ is a positive integer, and for each $1 \leq i \leq m$, $r_i$ is a non-empty string and $a_i$ and $b_i$ are nonnegative integers such that $a_i + b_i > 0$. For example, $\prod_{n \geq 0} \mathtt{a}^{n+1}\mathtt{b} = \mathtt{abaabaaab}\cdots$ is a multilinear word. The class of multilinear words appears in [21] and also in [8] (as the reducts of the "prime" stream $\Pi$). Clearly the multilinear words properly include the ultimately periodic words. Any multilinear word which is not ultimately periodic we call **properly multilinear**.

### 2.2  Predictors

We now define predictors based on various types of automata. (See [23] for results on the original automata, which are language recognizers rather than predictors.) Each predictor $M$ takes as input an infinite word $\alpha$ and produces as output an infinite word $M(\alpha)$, with the restriction that for each $i \geq 1$, $M$ must output the $i$th symbol of $M(\alpha)$ before it can read beyond the $i-1$th symbol of $\alpha$. We call $M(\alpha)[i]$ $M$'s **guess** about position $i$ of $\alpha$. If $M(\alpha)[i] = \alpha[i]$ then we say that the guess is correct; otherwise we say that it is incorrect. If there is an $n \geq 1$ such that for every $i \geq n$, $M(\alpha)[i] = \alpha[i]$, then we say that $M$ **masters** $\alpha$. (If $M$ outputs only a finite number of symbols when given $\alpha$, then we say that $M(\alpha)$ is undefined and $M$ does not master $\alpha$.)

**DFA predictors**  A **DFA predictor** is a tuple $M = (Q, A, T, \triangleright, q_s)$, where $Q$ is the set of states, $A$ is the input alphabet, $\triangleright$ is the start-of-input marker, $q_s \in Q$ is the initial state, and $T$ is a transition function of the form $[Q \times (A \cup \{\triangleright\})] \to [Q \times A]$.

To perform a computation, $M$ is given an input consisting of the symbol $\triangleright$ followed by an infinite word $\alpha$. $M$ starts in state $q_s$ with its input head positioned at $\triangleright$. $M$ then makes transitions based on its current state and input symbol. At each transition, $M$ changes state, moves its head to the right, and makes a guess about what the next symbol will be. The sequence of these guesses constitutes $M(\alpha)$. More formally, let $C = [C_1, C_2, C_3, \ldots]$ where $C_i = \{[q_i, c_i, g_i]$ with $q_i \in Q$, $c_i \in (A \cup \{\triangleright\})$, $g_i \in A$ such that $q_1 = q_s$ and for each $i \geq 1$, $c_i = (\triangleright\alpha)[i]$ and $T(q_i, c_i) = [q_{i+1}, g_i]$. Notice that there is only one possible $C$, given $M$ and $\alpha$. Now for $i \geq 1$, set $M(\alpha)[i] = g_i$.

**DPDA predictors**  A **DPDA predictor** is a tuple $M = (Q, A, F, T, \triangleright, \triangle, q_s)$, where $Q$ is the set of states, $A$ is the input alphabet, $F$ is the stack alphabet, $\triangleright$ is the start-of-input marker, $\triangle$ is the bottom-of-stack marker, $q_s \in Q$ is the initial state, and $T$ is a transition function of the form

$$[Q \times (A \cup \{\triangleright\}) \times (F \cup \{\triangle\})] \to [Q \times (A \cup \{\mathsf{stay}\}) \times (F \cup \{\mathsf{pop}, \mathsf{keep}\})].$$

To perform a computation, $M$ is given an input consisting of the symbol $\triangleright$ followed by an infinite word $\alpha$. $M$ starts in state $q_s$ with stack $\triangle$ and with its input head positioned at $\triangleright$. $M$ then makes transitions based on its current state, input symbol, and stack symbol. At each transition, $M$ (1) changes state, (2) either moves its input head to the right and guesses what the next symbol will be, or else keeps it in place (using $\mathsf{stay}$), and (3) either pushes a symbol to the stack, pops the stack, or leaves it alone (using $\mathsf{keep}$). It is illegal for $M$ to pop $\triangle$. The sequence of guesses made by $M$ constitutes $M(\alpha)$.

**DSA predictors** A **DSA predictor** is a tuple $M = (Q, A, F, T, \triangleright, \triangle, q_s)$, where $Q$ is the set of states, $A$ is the input alphabet, $F$ is the stack alphabet, $\triangleright$ is the start-of-input marker, $\triangle$ is the bottom-of-stack marker, $q_s \in Q$ is the initial state, and $T$ is a transition function of the form

$$[Q \times (A \cup \{\triangleright\}) \times (F \cup \{\triangle\}) \times \{\mathsf{top}, \mathsf{inside}\}] \to$$
$$[Q \times (A \cup \{\mathsf{stay}\}) \times (F \cup \{\mathsf{pop}, \mathsf{keep}, \mathsf{up}, \mathsf{down}\})].$$

To perform a computation, $M$ is given an input consisting of the symbol $\triangleright$ followed by an infinite word $\alpha$. $M$ starts in state $q_s$ with stack $\triangle$ and with its input head positioned at $\triangleright$. $M$ then makes transitions based on its current state, input symbol, stack symbol, and whether or not the stack head is at the top of the stack ($\mathsf{top}$ means the stack head is at the top; $\mathsf{inside}$ means it is not). At each transition, $M$ (1) changes state, (2) either moves its input head to the right and guesses what the next symbol will be, or else keeps it in place (using $\mathsf{stay}$), and (3) either pushes a symbol to the stack, pops the stack, leaves it alone (using $\mathsf{keep}$), or moves its stack head up or down. It is illegal for $M$ to push or pop the stack when the stack head is not at the top of the stack, or to move it up when it is already at the top or down when it is already at the bottom. The sequence of guesses made by $M$ constitutes $M(\alpha)$.

**Multi-DFA predictors** A **multi-DFA predictor** is a tuple of the form $M = (Q, A, k, T, \triangleright, q_s)$, where $Q$ is the set of states, $A$ is the input alphabet, $k \geq 1$ is the number of input heads, $\triangleright$ is the start-of-input marker, $q_s \in Q$ is the initial state, and $T$ is a transition function of the form

$$[Q \times (A \cup \{\triangleright\})^k] \to [Q \times \{\mathsf{stay}, \mathsf{right}\}^k \times A].$$

To perform a computation, $M$ is given an input consisting of the symbol $\triangleright$ followed by an infinite word $\alpha$. $M$ starts in state $q_s$ with its $k$ input heads all positioned at $\triangleright$. $M$ then makes transitions based on its current state and the input symbols it sees under each of its heads. At each transition, $M$ (1) changes state, (2) for each head either moves it to the right or keeps it in place (using stay), and (3) makes a guess about what the next symbol will be. If in a given transition, $M$ does not reach a new input position (one which had not previously been reached by any head), $M$'s guess at that transition is disregarded (i.e., it is not included in $M(\alpha)$). That is, $M(\alpha)[i]$ is the guess of the first transition which moves any head to $\alpha[i]$.

A **sensing multi-DFA predictor** is a multi-DFA predictor extended so that its transition function takes an additional argument indicating, for each pair of heads, whether those two heads are at the same input position.

## 3 Prediction of periodic words

In this section we study finite automata, pushdown automata, stack automata, and multihead finite automata as predictors of purely periodic and ultimately periodic words.

### 3.1 Prediction by DFAs

**Theorem 1.** *Let $A$ be an alphabet such that $|A| \geq 2$. Then no DFA predictor masters every purely periodic word over $A$.*

*Proof.* Suppose some DFA predictor $M$ masters every purely periodic word over $A$. $M$ has some number of states $p$. Take any $a, b \in A$ such that $a \neq b$. Let $\alpha$ be the purely periodic word $(a^{p+1}b)^\omega$. Then there is an $n \geq 1$ such that for every $i \geq n$, $M(\alpha)[i] = \alpha[i]$. Take the first segment of $p+1$ consecutive $a$s after the position $n$. At two of these $a$s, $M$ is in the same state. Then $M$ will repeat the guesses it made between those two $a$s for as long as it keeps reading $a$s. But then $M$ will guess $a$ for the next $b$, a contradiction. So $M$ does not master $\alpha$. □

### 3.2 Prediction by DPDAs

**Theorem 2.** *Let $A$ be an alphabet such that $|A| \geq 2$. Then no DPDA predictor masters every purely periodic word over $A$.*

*Proof (Sketch).* Suppose some DPDA predictor $M = (Q, A, F, T, \triangleright, \triangle, q_s)$ masters every purely periodic word over $A$. We set $p$ to be very large with respect to $|Q|$ and $|F|$. Take any $a, b \in A$ such that $a \neq b$. Let $\alpha$ be the purely periodic word $(a^p b)^\omega$. Then there is some position $m \geq 0$ after which all of $M$'s guesses about $\alpha$ are correct. Now, between each two segments of $p$ consecutive $a$'s, there is only one symbol (a single $b$), so the stack can grow by at most $|Q| \cdot |F|$ between each two segments. It follows that in some segment of $p$ consecutive $a$'s occurring after $m$, the stack height does not decrease by more than $|Q| \cdot |F|$, since otherwise it would eventually become negative. We show that in such a segment, because $p$ is so large with respect to $|Q|$ and $|F|$, there are two configurations $C_i$ and $C_j$ of $M$ occurring at different input positions with the same state and stack symbol, such that the stack below the top symbol at $C_i$ is not accessed between $C_i$ and $C_j$. Then since all of $M$'s guesses between $C_i$ and $C_j$ are $a$'s, $M$ will continue to guess $a$'s for as long as it continues to read $a$'s. But then $M$ will guess $a$ for the $b$ at the end of the segment, contradicting the supposition that all of $M$'s guesses about $\alpha$ after $m$ are correct. Therefore $M$ does not master every purely periodic word over $A$. □

### 3.3 Prediction by DSAs

We give two results about the predictive capabilities of DSAs: first, that some DSA predictor masters every purely periodic word, and second, that no DSA predictor can master any infinite word which is not multilinear.

---

**Algorithm 1** A DSA predictor which masters every purely periodic word. The input head is denoted by $h_i$ and the stack head is denoted by $h_s$. The input consists of the symbol ▷ followed by an infinite word $\alpha$. Wherever a guess is not specified, it may be taken to be arbitrary.

---

```
 1: loop
 2:     move h_i
 3:     push α[h_i]
 4:     recovering ← false
 5:     loop
 6:         move h_s down until stack[h_s] = △
 7:         matched ← true
 8:         loop
 9:             move h_s up
10:             move h_i, guessing stack[h_s]
11:             matched ← false if α[h_i] ≠ stack[h_s]
12:             break if top
13:         recovering ← true if not matched
14:         break if recovering and matched
```

---

**Theorem 3.** *Let $A$ be an alphabet. Then some DSA predictor masters every purely periodic word over $A$.*

*Proof.* Let $M$ be a DSA predictor which implements Algorithm 1. (The boolean variables *recovering* and *matched* can be accommodated using $M$'s finite state control.) The idea is that $M$ will gradually build up its stack until the stack consists of the period (or a cyclic shift thereof) of the purely periodic word to be mastered. Following Algorithm 1, $M$ begins by pushing the first symbol of the input after ▷ onto its stack, and then enters the loop spanning lines 5–14. This loop moves the stack head to the bottom of the stack and then moves it up symbol by symbol, predicting that the input will match the stack. Call each iteration of the loop spanning lines 5–14 a "pass", and call a pass successful if *matched* is true at line 14 and unsuccessful otherwise. Observe that if a pass is successful, then all of the guesses made during it (on line 10) are correct, and that if eventually there are no more unsuccessful passes, then $M$ masters its input.

Now take any purely periodic word $\alpha = x^\omega$. To show that $M$ masters $\alpha$, we first show that every unsuccessful pass will eventually be followed by a successful pass. Observe that there must be at least one successful pass, since $M$ begins the passes with only one symbol on the stack, and that symbol will eventually reappear in the input. So take any unsuccessful pass after the first successful pass. Now take the most recent successful pass prior to that unsuccessful pass. Let $i$ be the position of the input head in $x$ (counting from zero, so $0 \leq i < |x|$) at the beginning of this most recent successful pass and let $h$ be the height of the stack. Then the position of the input head in $x$ after the successful pass is $(i + h) \bmod |x|$. Then after $|x| - 1$ unsuccessful passes, the position of the input head in $x$ will be $(i + h|x|) \bmod |x| = i$. So the next pass after that will be successful. Hence every unsuccessful pass will eventually be followed by a successful pass.

Since each unsuccessful pass sets *recovering* to true, the next successful pass after it will break at line 14, causing $M$ to push another symbol onto the stack. If the height of the stack never reaches $|x|$, then after some point, every pass is successful and $M$ masters $\alpha$. So say the height of the stack eventually reaches $|x|$. Then since the last pass before the stack reached that height was successful, and the input symbol following that pass is now at the top of the stack, the previous $|x|$ symbols of the input match the stack. Then every subsequent pass will be successful, and $M$ masters $\alpha$. □

**Theorem 4.** *Every infinite word mastered by a DSA predictor is multilinear.*

*Proof.* Let $M$ be a DSA predictor and let $\alpha$ be any infinite word mastered by $M$. We will show that there is a DSA recognizer for $\text{Prefix}(\alpha)$, the set of all prefixes of $\alpha$. Since $M$ masters $\alpha$, there is an $n \geq 1$ such that for every $i \geq n$, $M(\alpha)[i] = \alpha[i]$. Take any such $n$. Let $C = (q, s, i)$ be the configuration of $M$ upon reaching position $n$ of $\alpha$, where $q$ is the state of $M$, $s$ is the stack, and $i$ is the position of the stack head within $s$. Let $M_\alpha$ be a DSA recognizer which operates as follows. First $M_\alpha$ uses its finite control to check that the first $n$ symbols of its input match the first $n$ symbols of $\alpha$. Then $M_\alpha$ uses its finite control to push $s$ onto its stack and move its stack head to position $i$ within $s$. Next $M_\alpha$ simulates $M$, starting from $C$. Whenever $M$ would make a guess, $M_\alpha$ instead checks that the next symbol of the input matches $M$'s guess. If any check fails, then $M_\alpha$ rejects its input; otherwise, when $M_\alpha$ reaches end-of-input, it accepts. Since all of $M$'s guesses after $n$ are correct, $M_\alpha$ now recognizes $\text{Prefix}(\alpha)$, and hence $M_\alpha$ determines $\alpha$ in the sense of [21]. Then by Theorem 8 of [21], $\alpha$ is multilinear. □

### 3.4  Prediction by multi-DFAs

We next consider multi-DFA predictors. We leave their more powerful cousins, sensing multi-DFA predictors, to Section 4.

**Algorithm 2** A 2-head DFA predictor which masters every ultimately periodic word. The heads are denoted by $t$ and $h$. The input consists of the symbol ▷ followed by an infinite word $\alpha$. Wherever a guess is not specified, it may be taken to be arbitrary.

---

   move $h$
   **loop**
      move $t$
      move $h$, guessing $\alpha[t]$
      move $h$ **if** $\alpha[h] \neq \alpha[t]$

---

**Theorem 5.** *Let $A$ be an alphabet. Then some multi-DFA predictor masters every ultimately periodic word over $A$.*

*Proof.* We employ a variation of the "tortoise and hare" cycle detection algorithm [19], adapted to our setting. Let $M$ be a 2-head DFA predictor which implements Algorithm 2. Take any ultimately periodic word $\alpha = xy^\omega$. Following the algorithm, the two heads $t$ (for "tortoise") and $h$ (for "hare") begin at the start of the input. $M$ moves $h$ one square to the right (making an arbitrary guess) and then enters the loop. In the loop, $M$ guesses that $h$ will match $t$. After each missed guess, $h$ moves ahead an extra square (making an arbitrary guess), so the distance between the two heads increases by 1. If this distance stops growing, then there are no more missed guesses, so $M$ masters $\alpha$. Otherwise, both heads will reach the periodic part $y^\omega$ of $\alpha$ and the distance between them will reach a multiple of $|y|$. Then each head will point to the same position in $y$ as the other, so all guesses will be correct from that point on. So again $M$ masters $\alpha$. □

## 4  Prediction of multilinear words

We turn now to prediction of the class of multilinear words. We give an algorithm by which a sensing multi-DFA can master every multilinear word.

**Theorem 6.** *Let $A$ be an alphabet. Then some sensing multi-DFA predictor masters every multilinear word over $A$.*

*Proof (Sketch).* Let $M$ be a sensing 10-head DFA predictor which implements Algorithm 3. The idea of the algorithm is as follows. Any properly multilinear word $\alpha$ can be written as $q \prod_{n \geq 1} \prod_{i \geq 1}^{m} p_i s_i^n$ for some $m \geq 1$ and strings $q$, $p_i$, $s_i$ subject to certain conditions. That is, $\alpha$ can be broken into "blocks", each block consisting of $m$ "segments" of the form $p_i s_i^n$. To master $\alpha$, $M$ will alternate between two procedures, CORRECTION and MATCHING. CORRECTION attempts to position $h_1$, $h_2$, $h_3$, and $h_4$ so that each head is at the beginning of a segment, $h_2$ is ahead of $h_1$ by a given number of segments, $h_3$ is ahead of $h_2$ by the same number of segments, and $h_4$ is ahead of $h_3$ by the same number of segments. Each time CORRECTION is entered, the given number of segments used to separate the heads is increased by one. MATCHING attempts to master $\alpha$ on the assumption that CORRECTION has successfully positioned $h_1$, $h_2$, $h_3$, and $h_4$ at the beginning of segments and that the number of segments separating the heads is a multiple of $m$ (meaning that the segments share the same $p_i$ and $s_i$). If any problem is detected, MATCHING is exited and CORRECTION is entered again.

The number of segments used to separate the heads is given by $r - l$. Before each call to CORRECTION, $r$ is moved forward, increasing this number by one. CORRECTION works by first moving $h_1$ forward to $h_4$ and then calling ADVANCEONE(1), which tries to move $h_1$ to the beginning of the next segment. Then CORRECTION moves $h_2$ to $h_1$ and calls ADVANCEMANY(2), which tries to move $h_2$ forward by $r - l$ segments. CORRECTION then moves $h_3$ to $h_2$ and calls ADVANCEMANY(3), which tries to move $h_3$ forward by $r - l$ segments. Finally, CORRECTION moves $h_4$ to $h_3$ and calls ADVANCEMANY(4), which tries to move $h_4$ forward by $r - l$ segments. If everything worked as intended, the four heads are now at the beginning of segments and each pair of heads $h_i$ and $h_{i+1}$ are separated by the same number of segments, $r - l$.

MATCHING works by using $h_1$, $h_2$, and $h_3$ to predict $h_4$. If the four heads are separated by the same number of segments, and if this number is a multiple of $m$, then the heads share the same $p_i$ and $s_i$. In this case, the later heads have extra copies of $s_i$: for some $d \geq 1$, in each segment $i$, $h_4$ will see $d$ more copies of $s_i$ than $h_3$, which will see $d$ more than $h_2$, which will see $d$ more than $h_1$. MATCHING moves the heads together, using the earlier heads to predict $h_4$ and detecting when each head passes its last copy of $s_i$ by comparing the heads with each other. By use of a normal form for properly multilinear words, we guarantee that the first symbol of $p_{i+1}$ differs from the first symbol of $s_i$, ensuring that the next segment can be detected. The supplemental head $h_{3a}$ is used to predict $h_4$'s last $d$ copies of $s_i$ by using $h_3$'s last $d$ copies a second time. Once all heads are at the beginning of the next segment, MATCHING repeats from the start. If any guess is incorrect, then the heads were not separated by a multiple of $m$ segments when MATCHING was entered. Upon making an incorrect guess, MATCHING exits, $r - l$ is increased, and CORRECTION is entered again.

The fact that $M$ is sensing allows it to perform operations a designated number of times, a technique used in the procedures ADVANCEMANY and ADVANCEONE called by CORRECTION. This technique works in the following way. Let $n$ be the distance between the heads $l$ and $r$ at a given point in the computation. To perform an operation $n$ times, we first move another head, say *inner*, to $r$. Then we move $l$ and $r$ together until $l$ reaches *inner*, performing the operation after each step. Now the operation has been performed $n$ times, and we can repeat this process to perform it another $n$ times. Further, by increasing the distance between $l$ and $r$, we can increase $n$. It is also possible to nest this process, by moving another head, say *outer*, to $r$, keeping *outer*'s position constant relative to $l$ and $r$ during the inner process, and moving $l$ and $r$, but not *outer*, each time the inner process is completed. When $l$ reaches *outer*, the inner process has been executed $n$ times, each time performing its operation $n$ times. In ADVANCEMANY and ADVANCEONE, this technique is used to advance a given $h_i$ by $n$ segments, using within each segment a threshold based on $n$ to detect the beginning of the next segment.

---

**Algorithm 3** A sensing 10-head DFA predictor which masters every multilinear word. The heads are denoted by $h_1$, $h_2$, $h_{3a}$, $h_3$, $h_4$, $t$, $l$, $r$, $inner$, and $outer$. The input consists of the symbol $\triangleright$ followed by an infinite word $\alpha$. Wherever a guess is not specified, it may be taken to be arbitrary.

---

**loop**
    move $r$
    CORRECTION
    MATCHING

**procedure** MATCHING
    **loop**
        move $h_{3a}$ **until** $h_{3a} = h_3$

        **while** $\alpha[h_1] = \alpha[h_2] = \alpha[h_3] = \alpha[h_4]$ **do**
            move $h_1, h_2, h_{3a}, h_3$
            move $h_4$, guessing $\alpha[h_2]$
        **break unless** $\alpha[h_2] = \alpha[h_4]$

        **while** $\alpha[h_2] = \alpha[h_3] = \alpha[h_4]$ **do**
            move $h_2, h_3$
            move $h_4$, guessing $\alpha[h_3]$
        **break unless** $\alpha[h_3] = \alpha[h_4]$

        **while** $\alpha[h_{3a}] = \alpha[h_3] = \alpha[h_4]$ **do**
            move $h_{3a}, h_3$
            move $h_4$, guessing $\alpha[h_{3a}]$
        **break unless** $\alpha[h_{3a}] = \alpha[h_4]$

        **while** $h_{3a} \neq h_3$ **and** $\alpha[h_{3a}] = \alpha[h_4]$ **do**
            move $h_{3a}$
            move $h_4$, guessing $\alpha[h_{3a}]$
        **break unless** $\alpha[h_{3a}] = \alpha[h_4]$

**procedure** CORRECTION
    move $h_1$ **until** $h_1 = h_4$
    ADVANCEONE(1)

    move $h_2$ **until** $h_2 = h_1$
    ADVANCEMANY(2)

    move $h_3$ **until** $h_3 = h_2$
    ADVANCEMANY(3)

    move $h_4$ **until** $h_4 = h_3$
    ADVANCEMANY(4)

**procedure** ADVANCEMANY($i$)
    move $outer$ **until** $outer = r$
    **while** $l \neq outer$ **do**
        ADVANCEONE($i$)
        move $l, r$

**procedure** ADVANCEONE($i$)
    move $t$ **until** $t = h_i$
    move $h_i$
    move $inner$ **until** $inner = r$
    **while** $l \neq inner$ **do**
        **if** $\alpha[t] = \alpha[h_i]$ **then**
            move $l, r, outer$
        **else**
            move $inner$ **until** $inner = r$
            move $h_i$
        move $t$
        move $h_i$
    **while** $\alpha[t] = \alpha[h_i]$ **do**
        move $t$
        move $h_i$, guessing $\alpha[t]$

---

To show that $M$ masters every multilinear word $\alpha$, we first show that if either MATCHING or CORRECTION gets "stuck", i.e. is entered and does not end, then in its stuck state it will continue to make guesses, all of which are correct, and so $M$ masters $\alpha$. In particular, we show that the first **while** loop of ADVANCEONE will always end. This loop implements the "tortoise and hare" routine of Algorithm 2 on $\alpha$, waiting for a streak of $r - l$ consecutive matches. Such a streak will eventually be obtained, because if $\alpha$ is ultimately periodic, then by the proof of Theorem 5, the "tortoise and hare" algorithm masters $\alpha$, and if $\alpha$ is properly multilinear, then we show that the "tortoise and hare" algorithm will eventually achieve $k$ consecutive matches on $\alpha$ for any $k \geq 1$, and so the loop will end.

So we are left with the case in which MATCHING and CORRECTION always end. Since $r$ is moved at the beginning of each iteration of the main loop, and since CORRECTION and MATCHING leave $r - l$ unchanged, $r - l$ will grow. If $\alpha$ is ultimately periodic, then eventually $r - l$ will be large enough for ADVANCEONE to "line up" the heads $h_i$ and $t$ with respect to the periodic part of $\alpha$, so that $M$ masters $\alpha$. If $\alpha$ is properly multilinear, then eventually $r - l$ will be large enough for ADVANCEONE to always advance $h_i$ by at least one segment. We show further that $r - l$ will grow slowly enough with respect to the segment length that eventually whenever $h_i$ is at the beginning of a segment, ADVANCEONE will move it to the beginning of the next segment and not farther. As a result, eventually CORRECTION will always end with the four heads $h_1$, $h_2$, $h_3$, and $h_4$ at the beginning of segments, with the heads separated by $r - l$ segments as desired. When $r - l$ next reaches a multiple of $m$, the segments of the four heads will share the same $p_i$ and $s_i$. We show that then MATCHING can make use of $h_1$, $h_2$, and $h_3$ to correctly predict $h_4$ as intended. Thus $M$ masters $\alpha$.   □

## 5  Conclusion

In this paper, we studied the classic problem of sequence prediction from the angle of automata and infinite words. We examined several types of automata and sought to find out which classes of infinite words they could master. In doing so we described novel prediction algorithms for the classes of purely periodic, ultimately periodic, and multilinear words. Open questions in our investigation include whether there is a DSA predictor which masters every ultimately periodic word, and whether there is a multi-DFA predictor without sensing which masters every multilinear word. Other directions for further research would be to consider other types of automata as predictors, e.g. automata with two-way input tapes, and to attempt prediction of other classes of infinite words, e.g. morphic words. It would also be interesting to consider questions of computational tractability, e.g. how many guesses and how much time is required to achieve mastery.

**Acknowledgments.** I would like to thank my Ph.D. advisor at Northeastern, Rajmohan Rajaraman, for his helpful comments and suggestions. The continuation of this work at Marne-la-Vallée was supported by the Agence Nationale de la Recherche (ANR) under the project EQINOCS (ANR-11-BS02-004).

## A  Prediction by DPDAs

In this appendix, we give a full proof of Theorem 2, filling out the sketch given in the body. This theorem states that no DPDA predictor masters every purely periodic word. We start with a lemma.

**Lemma 1.** *Take any integer $n \geq 1$ and let $L = m_1, \ldots, m_n$ be a list of integers such that for all $1 \leq i < n$, $|m_i - m_{i+1}| \leq 1$. Let $d = m_n - m_1$. Take any integer $k \geq 1$. Suppose $n \geq (2k-d)k^{2k-d}$. Then there are integers $1 \leq p_1 < \cdots < p_k \leq n$ such that for each $p_i$, for all $j$ such that $p_i \leq j \leq p_k$, $m_j \geq m_{p_i}$.*

*Proof.* Suppose there are $1 \leq a < b \leq n$ such that $m_b - m_a \geq k - 1$. Then for $1 \leq i \leq k$, set $p_i$ to the highest $j$ such that $j \leq b$ and $m_j = m_a + i - 1$. Then we are done.

So say there are no such $a, b$. Then we have $d < k$ and $m_n - k < m_i < m_1 + k$ for all $m_i$. Then there are at most $(m_1+k)-(m_n-k) = 2k-d$ distinct values in $L$. Then some value appears in $L$ at least $\frac{n}{2k-d} \geq k^{2k-d}$ times. For any integer $j$, let $|L|_j$ be the number of occurrences of $j$ in $L$. Take the lowest $m_n - k < j < m_1 + k$ such that $|L|_j \geq k^{j+k-m_n}$. If $j = m_n - k + 1$ then $j$ is the lowest value in $L$ and appears at least $k$ times, so choose $p_i$ from those appearances and we are done. Otherwise, $|L|_{j-1} < k^{j-1+k-m_n}$, so $|L|_j \geq k|L|_{j-1}$. Then there are $k$ appearances of $j$ in $L$ uninterrupted by $j-1$, so choose $p_i$ from those appearances and we are done. □

**Theorem 2.** *Let $A$ be an alphabet such that $|A| \geq 2$. Then no DPDA predictor masters every purely periodic word over $A$.*

*Proof.* Let $M = (Q, A, F, T, \triangleright, \triangle, q_s)$ be a DPDA predictor. Suppose $M$ masters every purely periodic word over $A$. Let $k = |Q| \cdot |F| + 1$ and let $p = (3k)k^{3k}$. Take any $a, b \in A$ such that $a \neq b$. Let $\alpha$ be the purely periodic word $(a^p b)^\omega$. Then there is some position $m \geq 0$ after which all of $M$'s guesses about $\alpha$ are correct.

Now, if the stack height increased by more than $|Q|\cdot|F|$ at one input position, there would be two configurations $C_1$ and $C_2$ of $M$ at that position with the same state and stack symbol, with $C_1$ occurring prior to $C_2$, such that the stack below the top symbol at $C_1$ is not accessed between $C_1$ and $C_2$. Then $M$ would loop and never reach the next input position. So the most that the stack height can increase at one position is $|Q|\cdot|F|$.

Let the stack difference of a segment of $p$ consecutive $a$'s be the height of the stack at the end of the segment minus the height of the stack at the beginning of the segment. Because there is only one symbol between each two segments (a single $b$), the stack height can increase by at most $|Q|\cdot|F|$ between segments. Then there must be a segment of $p$ consecutive $a$'s starting after position $m$ with a stack difference of at least $-|Q|\cdot|F|$, since otherwise the stack height after $m$ would eventually become negative.

So take any segment of $p$ consecutive $a$'s starting after $m$ with a stack difference $d \geq -|Q|\cdot|F|$. Let $C_1, \ldots, C_n$ be the successive configurations of $M$ during this segment, where each configuration $C_i$ has the form $(q_i, s_i)$, with $q_i$ being the current state and $s_i$ the current stack. We have $k \geq -d$ and $n \geq p$. Hence $n \geq (3k)k^{3k} \geq (2k-d)k^{2k-d}$. Then by Lemma 1 there is a list $P$ of integers $1 \leq p_1 < \cdots < p_k \leq n$ such that for each $p_i$, for all $j$ such that $p_i \leq j \leq p_k$, $|s_j| \geq |s_{p_i}|$. So since $k > |Q|\cdot|F|$, two of the $P$-indexed configurations $C_i$ and $C_j$ have the same state and stack symbol, with $i < j$. If $C_i$ and $C_j$ occurred at the same input position, then since the stack below the top symbol at $C_i$ is not accessed between $C_i$ and $C_j$, $M$ would loop and never reach the next input position. So $C_i$ and $C_j$ occur at distinct input positions $i_1 < i_2$ within the segment of $p$ consecutive $a$'s.

Now, all of the input symbols from $i_1$ to $i_2$ are $a$'s. Therefore as long as $M$ continues to read $a$'s it will repeat the computation between $i_1$ and $i_2$, since the stack below the top symbol at $i_1$ is not accessed between $i_1$ and $i_2$. So since all of $M$'s guesses from $i_1$ to $i_2$ are $a$'s, $M$ will continue to guess $a$'s for as long as it continues to read $a$'s. But then $M$ will guess $a$ for the $b$ at the end of the segment, contradicting the supposition that all of $M$'s guesses about $\alpha$ after $m$ are correct. Therefore $M$ does not master every purely periodic word over $A$. □

# B  Prediction of multilinear words

In this appendix, we give a full proof of Theorem 6, filling out the sketch given in the body. This theorem states that some sensing multi-DFA predictor masters every multilinear word. We begin by providing a normal form for properly multilinear words, together with some definitions to be used in the proofs. Then we prove a result about the behavior of Algorithm 2 ("tortoise and hare") when applied to multilinear words. With this groundwork laid, we show that by implementing Algorithm 3, a sensing multi-DFA can master every multilinear word.

### B.1   Normal form for properly multilinear words

In the theorem below we give a convenient form for properly multilinear words, resembling Proposition 32 of [8], but with a tighter constraint.

**Theorem 7.** *Let $\alpha$ be a properly multilinear word. Then $\alpha$ can be written as*

$$q \prod_{n \geq 1} \prod_{i \geq 1}^{m} p_i s_i^n$$

*for some $m \geq 1$, string $q$, and strings $p_i$ and $s_i$ such that*

- *for every $i$ from 1 to $m$, $p_i \neq \lambda$ and $s_i \neq \lambda$,*
- *for every $i$ from 1 to $m-1$, $s_i[1] \neq p_{i+1}[1]$, and*
- *$s_m[1] \neq p_1[1]$.*

*Proof.* By Theorem 15 of [21], $\alpha$ can be written as

$$q \prod_{n \geq 0} r_1^{a_1 n + b_1} r_2^{a_2 n + b_2} \cdots r_m^{a_m n + b_m}$$

for some $m \geq 1$, string $q$, non-empty strings $r_i$, and nonnegative integers $a_i$, $b_i$ where $a_i + b_i > 0$, such that

- for every $i$ from 1 to $m$, $b_i \geq 1$,
- for every $i$ from 1 to $m-1$, $r_i[1] \neq r_{i+1}[1]$, and
- if $m \geq 2$, $r_1[1] \neq r_m[1]$.

We transform this form into the desired one in four steps. Following [21], we view each $r_i^{a_i n + b_i}$ as a triple $[r_i, a_i, b_i]$. First, rotate the terms as described in Section 5 of [21] until $a_m$ is greater than 0. Second, split every triple $[r, a, b]$ such that $a > 0$ into two triples $[r^b, 0, 1]$ and $[r^a, 1, 0]$. Third, replace every triple $[r, 0, b]$ with $[r^b, 0, 1]$. Fourth, merge all adjacent triples $[r, 0, 1], [t, 0, 1]$ into $[rt, 0, 1]$ repeatedly until there are no more such adjacent triples. It is readily verified that the resulting list of triples consists of pairs $[p, 0, 1], [s, 1, 0]$ subject to the desired constraints. □

**Definitions for properly multilinear words** We now have that any properly multilinear word can be written as

$$q \prod_{n \geq 1} \prod_{i=1}^{m} p_i s_i^n$$

subject to the conditions of Theorem 7. In the context of a properly multilinear word $\alpha$ written subject to those conditions, we make the following definitions. Strings $p_i$ and $s_i$ are already defined for $1 \leq i \leq m$. Let $\rho = \max\{|p_i| \mid 1 \leq i \leq m\}$. Let $\sigma = \max\{|s_i| \mid 1 \leq i \leq m\}$. For each $n > m$, let $p_n = p_{((n-1) \mod m)+1}$ and $s_n = s_{((n-1) \mod m)+1}$. For each $n \geq 1$, let $block_n = \prod_{i=1}^{m} p_i s_i^n$ and $seg_n = p_n s_n^{\lceil \frac{n}{m} \rceil}$. We have $\alpha = q \prod_{n \geq 1} block_n = q \prod_{n \geq 1} seg_n$. For $j, k \geq 1$, we say that position $j$ of $\alpha$ occurs in **block** $k$ of $\alpha$, and write $block(j) = k$, iff $|q \prod_{n=1}^{k-1} block_n| < j \leq |q \prod_{n=1}^{k} block_n|$. (For $j \leq |q|$, we say that position $j$ does not occur in any block, and $block(j)$ is undefined.) For $j, k \geq 1$, we say that position $j$ of $\alpha$ occurs in **segment** $k$ of $\alpha$, and write $seg(j) = k$, iff $|q \prod_{n=1}^{k-1} seg_n| < j \leq |q \prod_{n=1}^{k} seg_n|$. (For $j \leq |q|$, we say that position $j$ does not occur in any segment, and $seg(j)$ is undefined.) Notice that for all $i > |q|$, $block(i) = \lceil \frac{seg(i)}{m} \rceil$.

### B.2  "Tortoise and hare" applied to multilinear words

In this subsection we show that if a multi-DFA predictor $M$ implements Algorithm 2 on a multilinear word, then for every $k \geq 1$, $M$ will at some point make $k$ consecutive correct guesses. We will make use of this result in the next subsection in proving that there is a sensing multi-DFA predictor which masters every multilinear word. We start with some lemmas.

**Lemma 2.** *Let $M$ be a multi-DFA predictor implementing Algorithm 2 on a properly multilinear word $\alpha$. Write $\alpha$ in the form of Theorem 7. Let $b = 2\rho + \sigma^2$. Suppose that while $h$ is in a segment $jh$ and $t$ is in a segment $jt$ such that $jh \mod m = jt \mod m$, $h$ moves $b$ symbols. Then $h$ and $t$ will agree afterward until $h$ leaves $jh$ or $t$ leaves $jt$, and if one leaves before the other, then at that point they will disagree.*

*Proof.* Consider the point at which $h$ begins to move the $b$ symbols. Since $jh \mod m = jt \mod m$, for some $1 \leq j \leq m$, segment $jh$ has the form $p_j s_j^{block(h)}$ and segment $jt$ has the form $p_j s_j^{block(t)}$. After $h$ moves $2|p_j|$ symbols, both heads are past $p_j$, so each head is inside some occurrence of $s_j$. Let $1 \leq dt \leq |s_j|$ be the position of $t$ within its occurrence of $s_j$ and let $1 \leq dh \leq |s_j|$ be the position of $h$ within its occurrence of $s_j$. Let $d = (dt - dh) \mod |s_j|$; $d$ indicates how many times $dh$ must be incremented with respect to $dt$ before $dh \mod |s_j| = dt \mod |s_j|$, at which point we say $h$ and $t$ have "lined up" with respect to $s_j$. Since $h$ is moved an extra symbol with respect to $t$ for each missed guess, if the two heads mismatch $d$ more times, they will be lined up. So after $h$ moves another $|s_j|^2$ symbols (making at most $b$ symbols in total), if the heads are not lined up, there were less than $d$ mismatches, hence at most $|s_j| - 2$ mismatches. Hence there were at least $|s_j|^2 - (|s_j| - 2) = |s_j|(|s_j| - 1) + 2$ guesses. Then by the pigeonhole principle, $M$ must have made $|s_j|$ consecutive correct guesses. So the heads are lined up or else $M$ has made $|s_j|$ consecutive correct guesses. Either way, since the same $|s_j|$ symbols will keep repeating under the two heads, $h$ and $t$ will now agree until $h$ leaves $jh$ or $t$ leaves $jt$. If one leaves before the other, then at that point they will disagree, since $s_j[1] \neq p_{j+1}[1]$. □

**Lemma 3.** *Let $M$ be a multi-DFA predictor implementing Algorithm 2 on a properly multilinear word $\alpha$. Write $\alpha$ in the form of Theorem 7. Suppose for some $k \geq 1$, $M$ never gets $k$ consecutive guesses correct. Then for every $d$, there is some point after which always $\operatorname{seg}(h) - \operatorname{seg}(t) \geq d$.*

*Proof.* Let $p = \sum_{i=1}^{m} |p_i|$ and $s = \sum_{i=1}^{m} |s_i|$. For each $n \geq 1$, let $\operatorname{sumblock1to}(n) = \sum_{i=1}^{n} |block_i|$. We have for all $n \geq 1$, $\operatorname{sumblock1to}(n) = \sum_{i=1}^{n} |block_i| = \sum_{i=1}^{n} p + is = np + \frac{ns(n+1)}{2}$. Now, since $M$ never gets $k$ consecutive guesses correct, and since $h$ moves an extra symbol ahead of $t$ on each missed guess, we have always $h \geq t + \frac{t}{k} - 2 \geq t(1 + \frac{1}{k}) - 2$. Now take any $b \geq 1$. Eventually $t$ will pass block $3bk$. Consider any point after that. There are $n \geq 3bk$ and $1 \leq c \leq |block_{n+1}|$ such that $t$ is on the $c$th symbol of block $n+1$. So $t = |q| + \operatorname{sumblock1to}(n) + c$ and $h \geq (|q| + \operatorname{sumblock1to}(n) + c)(1 + \frac{1}{k}) - 2$. We have

$$(1 + \frac{1}{k})\operatorname{sumblock1to}(n) = (1 + \frac{1}{k})(np + \frac{ns(n+1)}{2})$$
$$= (1 + \frac{1}{k})np + (1 + \frac{1}{k})n(n+1)\frac{s}{2}$$
$$= (1 + \frac{1}{k})np + ((1 + \frac{1}{k})n^2 + (1 + \frac{1}{k})n)\frac{s}{2}$$
$$= p(n + \frac{n}{k}) + (n^2 + \frac{n^2}{k} + n + \frac{n}{k})\frac{s}{2}$$
$$\geq p(n + 3b) + (n^2 + n + 3bn + 3b)\frac{s}{2}$$
$$> 2bp + p(n + b) + (n^2 + n + 2bn + bn + b)\frac{s}{2}$$
$$> 1 + p(n + b) + (n^2 + n + 2bn + b^2 + b)\frac{s}{2}$$
$$= 1 + p(n + b) + (n + b)(n + b + 1)\frac{s}{2}$$
$$= 1 + \operatorname{sumblock1to}(n + b),$$

giving us $(1 + \frac{1}{k})\operatorname{sumblock1to}(n) > \operatorname{sumblock1to}(n + b) + 1$. Then we have

$$h \geq (|q| + \operatorname{sumblock1to}(n) + c)(1 + \frac{1}{k}) - 2$$
$$= |q|(1 + \frac{1}{k}) + (1 + \frac{1}{k})\operatorname{sumblock1to}(n) + c(1 + \frac{1}{k}) - 2$$
$$> |q|(1 + \frac{1}{k}) + \operatorname{sumblock1to}(n + b) + 1 + c(1 + \frac{1}{k}) - 2$$
$$> |q| + \operatorname{sumblock1to}(n + b),$$

giving us $h > |q| + \text{sumblock1to}(n + b)$. Therefore $\text{block}(h) \geq n + b + 1$, so since $\text{block}(t) = n + 1$, we have $\text{block}(h) - \text{block}(t) \geq b$. So for every $b$, there is some point after which always $\text{block}(h) - \text{block}(t) \geq b$. So now take any $d$. Let $b = \frac{d-1}{m} + 1$. As shown above, there is some point after which always $\text{block}(h) - \text{block}(t) \geq b$. From that point onward, from the fact that each block contains exactly $m$ segments, we have always $\text{seg}(h) - \text{seg}(t) \geq m(b-1) + 1 \geq d$, which was to be shown. □

**Lemma 4.** *Let $M$ be a multi-DFA predictor implementing Algorithm 2 on a properly multilinear word $\alpha$. Write $\alpha$ in the form of Theorem 7. Suppose for some $k \geq 1$, $M$ never gets $k$ consecutive guesses correct. Then for every $n \geq 1$, there are segments $jh, jt \geq n$ of $\alpha$ such that $jh \mod m = jt \mod m$, $t$ enters $jt$ before $h$ enters $jh$, and $h$ leaves $jh$ before $t$ leaves $jt$.*

*Proof.* Take any $n \geq 1$. Take any segments $jh', jt' \geq n$ such that at some point, $h$ is in $jh'$ and $t$ is in $jt'$. Take any $d > jh' - jt'$ such that $d \mod m = 0$. By Lemma 3, there is some point after which always $\text{seg}(h) - \text{seg}(t) \geq d$. So there is a last point at which $\text{seg}(h) - \text{seg}(t) < d$. At this point, $h$ is in some segment $jh''$ and $t$ is in some segment $jt$ such that $jh'' - jt = d - 1$ and $jh'', jt \geq n$. If $t$ leaves $jt$ before $h$ leaves $jh''$, then $\text{seg}(h) - \text{seg}(t)$ would still be less than $d$, a contradiction. So $h$ leaves $jh''$ and enters $jh'' + 1$ before $t$ leaves $jt$. Now $\text{seg}(h) - \text{seg}(t) = d$. Now if $t$ leaves $jt$ before $h$ leaves $jh'' + 1$, then $\text{seg}(h) - \text{seg}(t)$ would again be less than $d$, a contradiction. So $h$ leaves $jh'' + 1$ before $t$ leaves $jt$. Letting $jh = jh'' + 1$, we therefore have that $t$ enters $jt$ before $h$ enters $jh$, and $h$ leaves $jh$ before $t$ leaves $jt$. Further, we have $jh, jt \geq n$, and since $jh - jt = d$ and $d \mod m = 0$, $jh \mod m = jt \mod m$, completing the proof. □

**Theorem 8.** *Let $M$ be a multi-DFA predictor implementing Algorithm 2 on a multilinear word $\alpha$. Then for every $k \geq 1$, $M$ will at some point make $k$ consecutive correct guesses.*

*Proof.* If $\alpha$ is ultimately periodic, then by the proof of Theorem 5, $M$ masters $\alpha$, so the statement holds. So say $\alpha$ is properly multilinear. Write $\alpha$ in the form of Theorem 7. Suppose for contradiction that for some $k \geq 1$, $M$ never gets $k$ consecutive guesses correct. Let $b = 2\rho + \sigma^2$. There is some $n \geq 1$ such that for every $n' \geq n$, $|seg_{n'}| \geq b + k$. Then by Lemma 4, there are segments $jh, jt \geq n$ such that $jh \mod m = jt \mod m$, $t$ enters $jt$ before $h$ enters $jh$, and $h$ leaves $jh$ before $t$ leaves $jt$. So $t$ is in $jt$ for the whole time that $h$ is in $jh$. Then by Lemma 2, once $h$ has moved $b$ symbols into $jh$, $h$ and $t$ will agree until $h$ reaches the beginning of segment $jh + 1$. Since $|seg_{jh}| \geq b + k$, $M$ therefore makes $k$ consecutive correct guesses, contradicting the supposition that $M$ never does so. So for every $k \geq 1$, $M$ will at some point make $k$ consecutive correct guesses. □

### B.3 Prediction of multilinear words by sensing multi-DFAs

We now give a full proof of Theorem 6, filling out the sketch given in the body. We prove lemmas about the matching and correction procedures, and then prove the main result. Algorithm 3 calls upon four procedures: MATCHING, CORRECTION, ADVANCEMANY, and ADVANCEONE. (The procedure ADVANCEONE takes a parameter $i \in \{1, 2, 3, 4\}$, and so is really four separate procedures; likewise for ADVANCEMANY.) All of the procedures have access to all of the heads of the predicting automaton. Below we prove lemmas about the behavior of these procedures when they are entered in certain "ready" configurations. Let $M$ be a sensing multi-DFA predictor with heads $h_1$, $h_2$, $h_3$, $h_{3a}$, $h_4$, $t$, $l$, $r$, $inner$, and $outer$, and let $\alpha$ be an infinite word. We say that $M$ is in a MATCHING-ready configuration on $\alpha$ if its heads are positioned on $\alpha$ such that $h_1 \leq h_2 \leq h_3 \leq h_4$ and $h_{3a} \leq h_3$. For each $1 \leq i \leq 4$, we say that $M$ is in an ADVANCE($i$)-ready configuration on $\alpha$ if its heads are positioned on $\alpha$ such that $t \leq h_i$, $l \leq r$, $inner \leq r$, and $outer \leq r$. We say that $M$ is in a CORRECTION-ready configuration on $\alpha$ if $M$ is in an ADVANCE(4)-ready configuration on $\alpha$ and $h_1 \leq h_2 \leq h_3 \leq h_4$.

**Matching procedure** We prove two lemmas about the matching procedure MATCHING.

**Lemma 5.** *Let $M$ be a sensing multi-DFA predictor in a MATCHING-ready configuration on an infinite word $\alpha$. If $M$ enters MATCHING and MATCHING does not end, then $M$ masters $\alpha$.*

*Proof.* MATCHING consists of an outer loop and four inner loops. If the first inner loop does not end, then $h_1$, $h_2$, $h_3$, and $h_4$ match, so guessing that $h_4$ matches $h_2$ is correct, and $M$ masters $\alpha$. If the second loop is entered and does not end, then $h_2$, $h_3$, and $h_4$ all match, so guessing that $h_4$ matches $h_3$ is correct, and $M$ masters $\alpha$. If the third loop does not end, then $h_{3a}$, $h_3$, and $h_4$ all match, so guessing that $h_4$ matches $h_{3a}$ is correct, and $M$ masters $\alpha$. If the fourth loop does not end, then $h_{3a}$ and $h_4$ match, so guessing that $h_4$ matches $h_{3a}$ is correct, and $M$ masters $\alpha$. So say the four inner loops always end. Now, each time the body of an inner loop is entered, at least one guess is made, and if any guess is missed in an inner loop, the outer loop ends immediately thereafter. So if the outer loop does not end and $M$ does not master $\alpha$, then at some point $M$ ceases entering the bodies of the inner loops. After that point, if the outer loop does not end immediately after skipping the first inner loop, then $h_2$ and $h_4$ match. Next, the second inner loop is skipped, so $h_2$, $h_3$, and $h_4$ do not all match, hence $h_3$ is different from $h_2$ and $h_4$. But then the outer loop ends immediately after the second inner loop. So MATCHING ends or $M$ masters $\alpha$. □

**Lemma 6.** *Let $M$ be a sensing multi-DFA predictor in a MATCHING-ready configuration on a properly multilinear word $\alpha$. Write $\alpha$ in the form of Theorem 7. Suppose that $h_1$, $h_2$, $h_3$, and $h_4$ are all at the beginning of segments, and for some $d \geq 1$, $\operatorname{seg}(h_2) - \operatorname{seg}(h_1) = \operatorname{seg}(h_3) - \operatorname{seg}(h_2) = \operatorname{seg}(h_4) - \operatorname{seg}(h_3) = dm$. Then if $M$ enters MATCHING, $M$ masters $\alpha$.*

*Proof.* We have that for some $i, j \geq 1$, for each $k \in \{1, 2, 3, 4\}$, $h_k$ is at the beginning of a string of the form $p_j s_j^{i+d(k-1)} p_{j+1}$. Recall that from Theorem 7, $s_j[1] \neq p_{j+1}[1]$. In MATCHING, $M$ first moves $h_{3a}$ to the same position as $h_3$. We depict the positions of the heads below. By **h** $s$ we mean that head $h$ is at the first symbol of string $s$.

```
··· h₁  pⱼ sⱼⁱ pⱼ₊₁ ···
··· h₂  pⱼ sⱼⁱsⱼᵈ pⱼ₊₁ ···
··· h₃ₐ h₃ pⱼ sⱼⁱsⱼᵈsⱼᵈ pⱼ₊₁ ···
··· h₄  pⱼ sⱼⁱsⱼᵈsⱼᵈsⱼᵈ pⱼ₊₁ ···
```

$\cdots \mathbf{h_1}\ p_j\ s_j^i\ p_{j+1}\cdots$
$\cdots \mathbf{h_2}\ p_j\ s_j^i s_j^d\ p_{j+1}\cdots$
$\cdots \mathbf{h_{3a}}\ \mathbf{h_3}\ p_j\ s_j^i s_j^d s_j^d\ p_{j+1}\cdots$
$\cdots \mathbf{h_4}\ p_j\ s_j^i s_j^d s_j^d s_j^d\ p_{j+1}\cdots$

Following MATCHING, $M$ moves the heads until they disagree, which will happen after $|p_j s_j^i|$ symbols, when $h_1$ reaches $p_{j+1}$. In doing so $M$ guesses $h_2$ for $h_4$, and since $h_2$ and $h_4$ do not disagree, all of the guesses will be correct.

$\cdots p_j\ s_j^i\ \mathbf{h_1}\ p_{j+1}\cdots$
$\cdots p_j\ s_j^i\ \mathbf{h_2}\ s_j^d\ p_{j+1}\cdots$
$\cdots p_j\ s_j^i\ \mathbf{h_{3a}}\ \mathbf{h_3}\ s_j^d s_j^d\ p_{j+1}\cdots$
$\cdots p_j\ s_j^i\ \mathbf{h_4}\ s_j^d s_j^d s_j^d\ p_{j+1}\cdots$

Next, $M$ moves $h_2$, $h_3$, and $h_4$ together until they disagree, which will happen after $|s_j^d|$ symbols, when $h_2$ reaches $p_{j+1}$. In doing so $M$ guesses $h_3$ for $h_4$, and since $h_3$ and $h_4$ do not disagree, all of the guesses will be correct.

$\cdots p_j\ s_j^i\ \mathbf{h_1}\ p_{j+1}\cdots$
$\cdots p_j\ s_j^i s_j^d\ \mathbf{h_2}\ p_{j+1}\cdots$
$\cdots p_j\ s_j^i\ \mathbf{h_{3a}}\ s_j^d\ \mathbf{h_3}\ s_j^d\ p_{j+1}\cdots$
$\cdots p_j\ s_j^i s_j^d\ \mathbf{h_4}\ s_j^d s_j^d\ p_{j+1}\cdots$

Next, $M$ moves $h_{3a}$, $h_3$, and $h_4$ together until they disagree, which will happen after $|s_j^d|$ symbols, when $h_3$ reaches $p_{j+1}$. In doing so $M$ guesses $h_{3a}$ for $h_4$, and since $h_{3a}$ and $h_4$ do not disagree, all of the guesses will be correct.

$\cdots p_j\ s_j^i\ \mathbf{h_1}\ p_{j+1}\cdots$
$\cdots p_j\ s_j^i s_j^d\ \mathbf{h_2}\ p_{j+1}\cdots$
$\cdots p_j\ s_j^i s_j^d\ \mathbf{h_{3a}}\ s_j^d\ \mathbf{h_3}\ p_{j+1}\cdots$
$\cdots p_j\ s_j^i s_j^d s_j^d\ \mathbf{h_4}\ s_j^d\ p_{j+1}\cdots$

Finally, $M$ moves $h_{3a}$ and $h_4$ together until $h_{3a}$ reaches $h_3$ or $h_{3a}$ and $h_4$ disagree. (Here $M$ uses its sensing ability to detect coincidence of $h_{3a}$ and $h_3$.) Since $h_{3a}$ and $h_4$ agree for the next $|s_j^d|$ symbols, $h_{3a}$ will reach $h_3$.

$$
\boxed{\begin{array}{l}
\cdots p_j\ s_j^i\ \mathbf{h_1}\ p_{j+1}\cdots \\
\cdots p_j\ s_j^i s_j^d\ \mathbf{h_2}\ p_{j+1}\cdots \\
\cdots p_j\ s_j^i s_j^d s_j^d\ \mathbf{h_{3a}}\ \mathbf{h_3}\ p_{j+1}\cdots \\
\cdots p_j\ s_j^i s_j^d s_j^d s_j^d\ \mathbf{h_4}\ p_{j+1}\cdots
\end{array}}
$$

Now all of the heads are at $p_{j+1}$, and the above process will repeat. Because no guesses were missed during this process, MATCHING will run perpetually without missing another guess, and so $M$ masters $\alpha$. □

**Correction procedure** The correction procedure consists of CORRECTION and its helper procedures ADVANCEONE and ADVANCEMANY. We give lemmas for these procedures first for ultimately periodic words, and then for properly multilinear words.

**Lemmas for the correction procedure (ultimately periodic case)**

**Lemma 7.** *Let $M$ be a sensing multi-DFA predictor in an* ADVANCE$(i)$-*ready configuration on an ultimately periodic word $\alpha$ for some $1 \leq i \leq 4$. Write $\alpha$ as $ps^\omega$ for strings $p, s$. If $M$ enters* ADVANCEONE$(i)$ *and* ADVANCEONE$(i)$ *does not end, then $M$ masters $\alpha$. Further, if $r - l \geq |ps|$ when* ADVANCEONE$(i)$ *is entered, then $M$ masters $\alpha$.*

*Proof.* When ADVANCEONE is entered, it moves $t$ until $t$ reaches $h_i$. At this point, $t$ and $h_i$ are at the beginning of an infinite word $\beta$, where $\alpha = \alpha[1..t]\beta$. Clearly the ultimately periodic words are closed under shifts, so $\beta$ is ultimately periodic. ADVANCEONE then implements the "tortoise and hare" routine of Algorithm 2 on $\beta$, waiting for a streak of $r - l$ consecutive matches of $t$ and $h_i$. By the proof of Theorem 5, this algorithm masters every ultimately periodic word, so such a streak will eventually be obtained. Finally, ADVANCEONE moves $t$ and $h_i$ together until they mismatch. If this happens, ADVANCEONE ends; if this never happens, then all of the guesses during this loop will be correct, so $M$ masters $\alpha$. If $r - l \geq |ps|$, the last $|s|$ guesses in the streak of $r - l$ consecutive correct guesses were made while both heads were past $p$. The last $|s|$ symbols of the streak will therefore keep repeating under both heads. So the two heads will continue to agree, and $M$ masters $\alpha$. □

**Lemma 8.** *Let $M$ be a sensing multi-DFA predictor in an* ADVANCE$(i)$-*ready configuration on an ultimately periodic word $\alpha$ for some $1 \leq i \leq 4$. If $M$ enters* ADVANCEMANY$(i)$ *and* ADVANCEMANY$(i)$ *does not end, then $M$ masters $\alpha$.*

*Proof.* ADVANCEMANY first moves $outer$ until $outer = r$, and then repeatedly calls ADVANCEONE on $h_i$ and moves $l$ and $r$ together. On each call to ADVANCEONE, by Lemma 7, ADVANCEONE will end, or $M$ masters $\alpha$. So if $M$ does not master $\alpha$, then after $r - l$ iterations of the loop, $l$ will catch up with $outer$, and ADVANCEMANY will end. □

**Lemma 9.** *Let $M$ be a sensing multi-DFA predictor in a* CORRECTION*-ready configuration on an ultimately periodic word $\alpha$. Write $\alpha$ as $ps^\omega$ for strings $p, s$. If $M$ enters* CORRECTION *and* CORRECTION *does not end, then $M$ masters $\alpha$. Further, if $r - l \geq |ps|$ when* CORRECTION *is entered, then $M$ masters $\alpha$.*

*Proof.* By Lemmas 7 and 8, each call to ADVANCEONE and ADVANCEMANY will end, or $M$ masters $\alpha$. So CORRECTION will end, or $M$ masters $\alpha$. If $r - l \geq |ps|$ when CORRECTION is entered, then $r - l \geq |ps|$ when ADVANCEONE is entered, so by Lemma 7, $M$ masters $\alpha$. □

**Lemmas for the correction procedure (properly multilinear case)**

**Lemma 10.** *Let $M$ be a sensing multi-DFA predictor in an* ADVANCE$(i)$*-ready configuration on a properly multilinear word $\alpha$ for some $1 \leq i \leq 4$. If $M$ enters* ADVANCEONE$(i)$, ADVANCEONE$(i)$ *will end, and it will move $h_i$ at least once.*

*Proof.* When ADVANCEONE is entered, it moves $t$ until $t$ reaches $h_i$. At this point, $t$ and $h_i$ are at the beginning of an infinite word $\beta$, where $\alpha = \alpha[1..t]\beta$. Clearly the properly multilinear words are closed under shifts, so $\beta$ is properly multilinear. ADVANCEONE then implements the "tortoise and hare" routine of Algorithm 2 on $\beta$, waiting for a streak of $r - l$ consecutive matches of $t$ and $h_i$. By Theorem 8, such a streak will eventually be obtained. Finally, ADVANCEONE moves $t$ and $h_i$ together until they mismatch, which must eventually happen, since $\beta$ is not ultimately periodic. So ADVANCEONE will end, and clearly it will have moved $h_i$ at least once. □

**Lemma 11.** *Let $M$ be a sensing multi-DFA predictor in an* ADVANCE$(i)$*-ready configuration on a properly multilinear word $\alpha$ for some $1 \leq i \leq 4$. Write $\alpha$ in the form of Theorem 7. Suppose $\rho + 2\sigma \leq r - l$. Then if $M$ enters* ADVANCEONE$(i)$ *with $h_i$ in some segment,* ADVANCEONE$(i)$ *will end with $h_i$ in a subsequent segment.*

*Proof.* Let $h = h_i$. When ADVANCEONE is entered, $h$ is in some segment $j$, so $h$ is at the beginning of a string of the form $ws_j^n p_{j+1}$, where $|w| \leq \max(|p_j|, |s_j|)$ and $0 \leq n \leq \text{block}(h)$. Suppose ADVANCEONE ends before $h$ reaches $p_{j+1}$. Since the required streak is $r - l$, $h$ and $t$ must each have moved at least $r - l$ symbols. Then since $r - l \geq |p_j| + 2|s_j|$, we have $n \geq 2$, and $h$ and $t$ are both in the $s_j^n$ part of segment $j$, past the first $s_j$. Let $c$ be the position of $t$ within $s_j$ and let $d$ be the position of $h$ within $s_j$. $t$ and $h$ agreed on the last $|s_j|$ symbols, so when $t$ was last at position $c$ within $s_j$, $h$ was at position $d$ within $s_j$, and $t$ and $h$ agreed on those positions. But then $s_j[c] = s_j[d]$, so $t$ and $h$ agree now, a contradiction, since they must disagree for ADVANCEONE to end. Therefore ADVANCEONE will not end before $h$ reaches $p_{j+1}$. But by Lemma 10, ADVANCEONE will end. So ADVANCEONE will end with $h$ in a subsequent segment. □

**Lemma 12.** *Let $M$ be a sensing multi-DFA predictor in an* ADVANCE$(i)$*-ready configuration on a properly multilinear word $\alpha$ for some $1 \leq i \leq 4$. Write $\alpha$ in* the form of Theorem 7. Suppose that $M$ enters ADVANCEONE($i$) with $h_i$ at the beginning of some segment $j$, and that $\rho + 2\sigma \leq r - l \leq |seg_j| - 2\rho - \sigma^2$. Then ADVANCEONE($i$) will end with $h_i$ at the beginning of segment $j + 1$.

*Proof.* Let $h = h_i$. By Lemma 11, ADVANCEONE will not end before $h$ reaches segment $j + 1$. Now when ADVANCEONE is entered, it moves $t$ until $t$ reaches $h$ and then implements the "tortoise and hare" routine of Algorithm 2, waiting for a streak of $r - l$ consecutive matches of $t$ and $h$. Let $b = 2\rho + \sigma^2$. Then by Lemma 2, once $h$ has moved $b$ symbols into segment $j$, $h$ and $t$ will agree until $h$ reaches the beginning of segment $j + 1$, at which point they will disagree. So since $|seg_j| \geq b + r - l$, $h$ and $t$ will achieve a streak of $r - l$ consecutive matches while in segment $j$. Then ADVANCEONE will enter the second while loop and move $t$ and $h$ together until they mismatch, which will happen when $h$ reaches the beginning of segment $j + 1$. □

**Lemma 13.** *Let $M$ be a sensing multi-DFA predictor in an ADVANCE($i$)-ready configuration on a properly multilinear word $\alpha$ for some $1 \leq i \leq 4$. Write $\alpha$ in the form of Theorem 7. Suppose that $M$ enters ADVANCEONE($i$) with $h_i$ in some segment $j$, and that $4(\rho + \sigma) \leq r - l \leq seg_{j+1} - \sigma^2 - 4(\rho + \sigma)$. Then ADVANCEONE($i$) will end with $h_i$ at the beginning of segment $j + 1$ or the beginning of segment $j + 2$.*

*Proof.* Let $h = h_i$. When ADVANCEONE is entered, it moves $t$ until $t$ reaches $h$ and then implements the "tortoise and hare" routine of Algorithm 2, waiting for a streak of $r - l$ consecutive matches of $t$ and $h$. Initially, $h$ is at the beginning of a string of the form $ws_j^{n'}p_{j+1}s_{j+1}^{n''}p_{j+2}$ where $|w| \leq max(|p_j|, |s_j|)$, $n' \geq 0$, and $n'' = \text{block}(h)$ or $\text{block}(h) + 1$. By Lemma 11, ADVANCEONE will not end with $h$ in segment $j$. So $h$ will reach the beginning of $p_{j+1}$. If ADVANCEONE ends now, then the lemma is satisfied. So say ADVANCEONE does not end at this point.

Then consider the situation with $h$ at the beginning of the string $p_{j+1}s_{j+1}^{n''}p_{j+2}$. We have $h - t \leq |w| + |s_j|$, since if $t$ has not reached $s_j^{n'}$, then $h - t \leq |w|$, and if $t$ reached $s_j^{n'}$, then $h$ was at most $|w|$ ahead of it, and with both of them in $s_j^{n'}$, they could separate by at most another $|s_j|$ before reaching identical positions in $s_j$, after which they would not separate further. Now let $s$ be the current streak. Suppose $s > |p_j| + 2|s_j|$. Then since $t$ has moved at least $s$ symbols, $t$ is in $s_j^{n'}$, past the first $s_j$. Let $c$ be the position of $t$ within $s_j$. $t$ and $h$ agreed on the last $|s_j|$ symbols, so when $t$ was last at position $c$ within $s_j$, $h$ was at position 1 within $s_j$, since now $h$ is at a position following the last position of $s_j$. $t$ and $h$ agreed on those positions, so $s_j[c] = s_j[1]$. But since the streak was not reset when $h$ reached $p_{j+1}$, $t$ and $h$ are still in agreement, so $s_j[c] = p_{j+1}[1]$, giving $s_j[1] = p_{j+1}[1]$, a contradiction. So $s \leq |p_j| + 2|s_j|$.

Now, $t$ is at most $|w| + |s_j|$ symbols behind $h$, and therefore at most $|w| + |s_j| + |p_{j+1}|$ symbols behind the start of $s_{j+1}^{n''}$. $t$ will reach the start of the second $s_{j+1}$, since at that point the streak is at most $|p_j| + 2|s_j| + |w| + |s_j| + 2|p_{j+1}|$, which is less than $r - l$. Then the procedure will not end before $h$ reaches $p_{j+2}$, since if it did, $t$ and $h$ would disagree while both in $s_{j+1}^{n''}$, after an $|s_{j+1}|$ streak with both in $s_{j+1}^{n''}$, which is impossible. So given that AdvanceOne did not end with $h$ at the beginning of $p_{j+1}$, $h$ will reach the beginning of $p_{j+2}$.

Now when $h$ reached $p_{j+1}$, $t$ was at most $|w|+|s_j|$ symbols behind $h$, so when $t$ reaches $p_{j+1}$, $t$ is at most $2(|w|+|s_j|)$ symbols behind $h$. Let $b = 2\rho + \sigma^2$. Then the number of symbols remaining ahead of $h$ in segment $j+1$ is at least

$$|seg_{j+1}| - 2(|w|+|s_j|)$$
$$\geq r - l + \sigma^2 + 4(\rho + \sigma) - 2(|w|+|s_j|)$$
$$= r - l + b + 2\rho + 4\sigma - 2(|w|+|s_j|)$$
$$\geq r - l + b.$$

So by Lemma 2, once $h$ has moved another $b$ symbols, $h$ and $t$ will agree until $h$ reaches $p_{j+2}$, at which point they will disagree. So $h$ and $t$ will achieve a streak of $r-l$ consecutive correct guesses while in segment $j+1$. Then AdvanceOne will enter the second while loop and move $t$ and $h$ together until they mismatch, which happens when $h$ reaches the beginning of segment $j+2$. □

**Lemma 14.** *Let $M$ be a sensing multi-DFA predictor in an* Advance($i$)*-ready configuration on a properly multilinear word $\alpha$ for some $1 \leq i \leq 4$. If $M$ enters* AdvanceMany($i$)*, then* AdvanceMany($i$) *will end, and it will move $h_i$ at least once.*

*Proof.* AdvanceMany first moves *outer* until *outer* $= r$, and then repeatedly calls AdvanceOne($i$) and moves $l$ and $r$ together. Since $r - l \geq 1$, there will be at least one call to AdvanceOne. On each call to AdvanceOne, by Lemma 10, AdvanceOne will end, and it will move $h_i$ at least once. So after $r-l$ iterations of the loop, $l$ will catch up with *outer*, AdvanceMany will end, and it will have moved $h_i$ at least once. □

**Lemma 15.** *Let $M$ be a sensing multi-DFA predictor in an* Advance($i$)*-ready configuration on a properly multilinear word $\alpha$ for some $1 \leq i \leq 4$. Write $\alpha$ in the form of Theorem 7. Suppose $\rho + 2\sigma \leq r - l$. Then if $M$ enters* AdvanceMany($i$) *with $h_i$ in some segment $j$,* AdvanceMany($i$) *will end with* $\text{seg}(h_i) \geq j + r - l$.

*Proof.* AdvanceMany first moves *outer* until *outer* $= r$, and then repeatedly calls AdvanceOne($i$) and moves $l$ and $r$ together. On the first call to AdvanceOne, by Lemma 11, $h_i$ will be advanced from its current segment to some subsequent segment. Since AdvanceOne leaves $r-l$ unchanged, the same will be true for each subsequent call to AdvanceOne. So after $r-l$ iterations of the loop, $l$ will catch up with *outer* and AdvanceMany will end with $\text{seg}(h_i) \geq j + r - l$. □

**Lemma 16.** *Let $M$ be a sensing multi-DFA predictor in an* Advance($i$)*-ready configuration on a properly multilinear word $\alpha$ for some $1 \leq i \leq 4$. Write $\alpha$ in the form of Theorem 7. Suppose that $M$ enters* AdvanceMany($i$) *with $h_i$ at the beginning of some segment $j$ and $\rho + 2\sigma \leq r - l \leq \text{block}(h_i) - 2\rho - \sigma^2$. Then* AdvanceMany($i$) *will end with $h_i$ at the beginning of segment $j + r - l$.*

*Proof.* Let $h = h_i$. We have $\text{seg}(h) = j$, so for every segment $k \geq j$, $|seg_k| = |p_k s_k^{\lceil \frac{k}{m} \rceil}| \geq \lceil \frac{k}{m} \rceil \geq \lceil \frac{\text{seg}(h)}{m} \rceil = \text{block}(h)$. Hence for every segment $k \geq j$, we have $|seg_k| \geq \text{block}(h) \geq r - l + 2\rho + \sigma^2$. Therefore we can make use of Lemma 12 whenever $h$ is at the beginning of segment $j$ or any subsequent segment. Now, AdvanceMany first moves *outer* until $outer = r$, and then repeatedly calls AdvanceOne($i$) and moves $l$ and $r$ together. When AdvanceOne is first called, $h$ is at the beginning of a segment, so by Lemma 12, AdvanceOne will end with $h$ at the beginning of the next segment. Since AdvanceOne leaves $r-l$ unchanged, the same will be true for each subsequent call to AdvanceOne. After $r-l$ iterations of the loop, $l$ will catch up with *outer* and AdvanceMany will end, leaving $h$ at the beginning of segment $j + r - l$. □

**Lemma 17.** *Let $M$ be a sensing multi-DFA predictor in a* Correction-*ready configuration on a properly multilinear word $\alpha$. If $M$ enters* Correction, *then* Correction *will end, and it will move $h_4$ at least once.*

*Proof.* By Lemmas 10 and 14, each call to AdvanceOne and AdvanceMany will end, and $h_4$ will be moved at least once. So Correction will end, and it will have moved $h_4$ at least once. □

**Lemma 18.** *Let $M$ be a sensing multi-DFA predictor in a* Correction-*ready configuration on a properly multilinear word $\alpha$. Write $\alpha$ in the form of Theorem 7. Suppose $\rho + 2\sigma \leq r - l$. Then if $M$ enters* Correction *with $h_4$ in some segment $j$,* Correction *will end with* $\text{seg}(h_4) \geq j + 3(r-l) + 1$.*

*Proof.* Correction begins by moving $h_1$ until $h_1 = h_4$, and then runs AdvanceOne(1). By Lemma 11, AdvanceOne(1) will end with $\text{seg}(h_1) \geq j+1$. Next, $h_2$ is moved until $h_2 = h_1$ and then AdvanceMany(2) is called. Since $r - l$ is unchanged, by Lemma 15, AdvanceMany(2) will end with $\text{seg}(h_2) \geq j + 1 + r - l$. Next, $h_3$ is moved until $h_3 = h_2$ and then AdvanceMany(3) is called. Again since $r - l$ is unchanged, by Lemma 15, AdvanceMany(3) will end with $\text{seg}(h_3) \geq j + 1 + 2(r - l)$. Finally, $h_4$ is moved until $h_4 = h_3$ and AdvanceMany(4) is called. Again since $r - l$ is unchanged, by Lemma 15, AdvanceMany(4) will end with $\text{seg}(h_4) \geq j+1+3(r-l)$, completing the proof. □

**Lemma 19.** *Let $M$ be a sensing multi-DFA predictor in a* Correction-*ready configuration on a properly multilinear word $\alpha$. Write $\alpha$ in the form of Theorem 7. Suppose that $M$ enters* Correction *with $h_4$ in some segment $j$ and $4(\rho+\sigma) \leq r - l \leq \text{block}(h_4) - \sigma^2 - 4(\rho + \sigma)$. Then* Correction *will end with $h_1$ at the beginning of some segment $i > j$, $h_2$ at the beginning of segment $i + r - l$, $h_3$ at the beginning of segment $i + 2(r - l)$, and $h_4$ at the beginning of segment $i + 3(r - l)$.*

*Proof.* Correction begins by moving $h_1$ until $h_1 = h_4$, and then runs AdvanceOne(1). Then $\text{seg}(h_1) = j$, so we have $|seg_{j+1}| = |p_{j+1} s_{j+1}^{\lceil \frac{j+1}{m} \rceil}| \geq \lceil \frac{j+1}{m} \rceil \geq \lceil \frac{\text{seg}(h_1)}{m} \rceil = \text{block}(h_1)$. Therefore $|seg_{j+1}| \geq \text{block}(h_1)$, and hence we can make use of Lemma 13. So by Lemma 13, AdvanceOne(1) will end with $h_1$ at the beginning of either the next segment or of the one after it. So now $h_1$ is at the beginning of some segment $i > j$. Next, $h_2$ is moved until $h_2 = h_1$. Now $h_2$ is at the beginning of segment $i$, so since $r - l$ is unchanged, by Lemma 16, ADVANCEMANY(2) will end with $h_2$ at the beginning of segment $i + r - l$. Next, $h_3$ is moved until $h_3 = h_2$. Now $h_3$ is at the beginning of segment $i + r - l$, so again since $r - l$ is unchanged, by Lemma 16, ADVANCEMANY(3) will end with $h_3$ at the beginning of segment $i + 2(r - l)$. Finally, $h_4$ is moved until $h_4 = h_3$. Now $h_4$ is at the beginning of segment $i + 2(r - l)$, so again since $r - l$ is unchanged, by Lemma 16, ADVANCEMANY(4) will end with $h_4$ at the beginning of segment $i + 3(r - l)$, completing the proof. □

**Main loop** With lemmas for the matching and correction procedures in place, we are ready to prove the main result. We first give a lemma to establish that the procedures will always be entered in the "ready" configurations defined above.

**Lemma 20.** *Let $M$ be a sensing multi-DFA predictor which implements Algorithm 3 on an infinite word $\alpha$. Then whenever $M$ enters* MATCHING, *it is in a* MATCHING-*ready configuration, whenever $M$ enters* ADVANCEONE($i$) *or* ADVANCEMANY($i$) *for any $1 \leq i \leq 4$, it is in an* ADVANCE($i$)-*ready configuration, and whenever $M$ enters* CORRECTION, *it is in a* CORRECTION-*ready configuration.*

*Proof.* Let us say that $M$ is **CM-ready** if it is in a configuration on $\alpha$ which is both CORRECTION-ready and MATCHING-ready. At the beginning of Algorithm 3, all the heads are at the beginning of the input, so $M$ is CM-ready. In the main loop, $M$ moves $r$, then calls CORRECTION, and then calls MATCHING. If $M$ is CM-ready when it moves $r$, then it remains CM-ready after moving $r$.

Now, suppose $M$ is CM-ready when it calls CORRECTION. CORRECTION first moves $h_1$ until it reaches $h_4$. Since $M$ is CM-ready, it is in an ADVANCE(4)-ready configuration on $\alpha$, so $t \leq h_4$. Hence now $t \leq h_1$, so $M$ is in an ADVANCE(1)-ready configuration on $\alpha$. Now $M$ enters ADVANCEONE(1). Notice that ADVANCEMANY($i$) and ADVANCEONE($i$) never move $t$ past $h_i$, $l$ past $r$, *inner* past $r$, or *outer* past $r$. So whenever $M$ enters these procedures in an ADVANCE($i$)-ready configuration, it remains in an ADVANCE($i$)-ready configuration upon exiting them. Next, CORRECTION moves $h_2$ until it reaches $h_1$. Since $t \leq h_1$, we have now $t \leq h_2$, so $M$ is in an ADVANCE(2)-ready configuration on $\alpha$ when it enters ADVANCEMANY(2). Next, CORRECTION moves $h_3$ until it reaches $h_2$. Since $t \leq h_2$, we have now $t \leq h_3$, so $M$ is in an ADVANCE(3)-ready configuration on $\alpha$ when it enters ADVANCEMANY(3). Finally, CORRECTION moves $h_4$ until it reaches $h_3$. Since $t \leq h_3$, we have now $t \leq h_4$, so $M$ is in an ADVANCE(4)-ready configuration on $\alpha$ when it enters ADVANCEMANY(4). So if $M$ is CM-ready when it enters CORRECTION, then it is again CM-ready upon exiting CORRECTION.

Finally, notice that MATCHING never moves $h_1$ past $h_2$, $h_2$ past $h_3$, $h_3$ past $h_4$, or $h_{3a}$ past $h_3$. So if $M$ is CM-ready when it enters MATCHING, then it is again CM-ready upon exiting MATCHING. So CORRECTION and MATCHING are only entered when $M$ is CM-ready, completing the proof. □

Prediction of Infinite Words with Automata     29

**Theorem 6.** *Let A be an alphabet. Then some sensing multi-DFA predictor masters every multilinear word over A.*

*Proof.* Let $M$ be a sensing 10-head DFA predictor which implements Algorithm 3. By Lemma 20, whenever $M$ enters MATCHING, it is in a MATCHING-ready configuration, whenever $M$ enters ADVANCEONE($i$) or ADVANCEMANY($i$) for any $1 \leq i \leq 4$, it is in an ADVANCE($i$)-ready configuration, and whenever $M$ enters CORRECTION, it is in a CORRECTION-ready configuration. We can therefore make use of the lemmas proved above for the matching and correction procedures. To see that $M$ masters every multilinear word, take any such word $\alpha$. Suppose for contradiction that $M$ does not master $\alpha$.

First, suppose $\alpha$ is ultimately periodic. Then $\alpha = ps^\omega$ for some strings $p, s$ such that $s \neq \lambda$. By Lemma 5, if MATCHING is entered and does not end, then $M$ masters $\alpha$. By Lemma 9, if CORRECTION is entered and does not end, then $M$ masters $\alpha$. So since we supposed that $M$ does not master $\alpha$, both procedures always end. Then since $r$ is moved at the beginning of each iteration of the loop, and since CORRECTION and MATCHING leave $r - l$ unchanged, eventually CORRECTION will be entered with $r - l \geq |ps|$. Then by Lemma 9, $M$ masters $\alpha$, contradicting the supposition that it does not. So $M$ masters $\alpha$.

So say $\alpha$ is properly multilinear. Then $\alpha$ can be written as

$$q \prod_{n \geq 0} \prod_{i \geq 1}^{m} p_i s_i^n$$

subject to the conditions of Theorem 7 and the definitions of Section B.1. By Lemma 5, if MATCHING is entered and does not end, then $M$ masters $\alpha$. So since we supposed that $M$ does not master $\alpha$, MATCHING always ends. Now by Lemma 17, each time CORRECTION is entered, it will end, and it will move $h_4$ at least once. For each $i \geq 1$, let point $i$ be the point of the computation during the $i$th iteration of the loop of Algorithm 3, after $r$ has been moved but before CORRECTION has been entered. Since $r$ is moved at the beginning of each iteration of the loop, and since CORRECTION and MATCHING leave $r - l$ unchanged, we have for all $i \geq 1$, at point $i$, $r - l = i$. Let $j = 4(\rho + \sigma) + |q|$. Then for all $i \geq j$, at point $i$, $r - l \geq 4(\rho + \sigma)$ and $h_4 > |q|$. For each $i \geq j$, denote by $f(i)$ the value of $\text{seg}(h_4)$ at point $i$. We have $f(j) \geq 1$ and by Lemma 18, for all $i > j$, $f(i) \geq f(i-1) + 3i + 1$. Solving the recurrence, we get for all $i \geq j$, $f(i) \geq \frac{(i-j+1)(3(i-j)+2)}{2} \geq \frac{(i-j)^2}{2}$. Then for all $i \geq j$, at point $i$, $\text{block}(h_4) \geq \frac{(i-j)^2}{2m}$. Let $k = 2m(2j + \sigma^2 + 4\rho + 4\sigma) + j$. For all $i \geq k$, at point $i$, we have

$$\begin{aligned}
block(h_4) &\geq \frac{(i-j)^2}{2m} \\
&\geq \frac{(i-j)(k-j)}{2m} \\
&= (i-j)(2j + \sigma^2 + 4\rho + 4\sigma) \\
&= 2ij + i\sigma^2 + 4i\rho + 4i\sigma - j(2j + \sigma^2 + 4\rho + 4\sigma)
\end{aligned}$$

$$
\begin{aligned}
&= ij + i\sigma^2 + 4i\rho + 4i\sigma + ij - j(2j + \sigma^2 + 4\rho + 4\sigma) \\
&\geq ij + i\sigma^2 + 4i\rho + 4i\sigma \\
&\geq i + \sigma^2 + 4(\rho + \sigma) \\
&= r - l + \sigma^2 + 4(\rho + \sigma).
\end{aligned}
$$

Then for all $i \geq k$, at point $i$, we have $4(\rho+\sigma) \leq r-l \leq block(h_4) - \sigma^2 - 4(\rho+\sigma)$. So we can make use of Lemma 19 at any point $i \geq k$. Since $r - l$ increases by 1 with each iteration of the loop, for some $i \geq k$, at point $i$, $r-l$ is a multiple of $m$. Take any such $i$. Then when CORRECTION is entered on the $i$th iteration of the loop, by Lemma 19, it will exit with $h_1$ at the beginning of some segment $d$, $h_2$ at the beginning of segment $d+r-l$, $h_3$ at the beginning of segment $d+2(r-l)$, and $h_4$ at the beginning of segment $d + 3(r - l)$. Then the number of segments between $h_1$ and $h_2$ equals the number of segments between $h_2$ and $h_3$ equals the number of segments between $h_3$ and $h_4$ equals $r - l$, which is a multiple of $m$. So in the next call to MATCHING, by Lemma 6, $M$ masters $\alpha$, contradicting the supposition that it does not. So $M$ masters $\alpha$. ☐